\documentclass[12pt]{article}
\usepackage{graphicx}
\usepackage{latexsym,amsmath,amsfonts,amssymb,cite}
\usepackage{bbm}
\newcommand {\beq}{\begin{eqnarray}}
\newcommand {\eeq}{\end{eqnarray}}

\def \be {\begin{equation}}
\def \ee {\end{equation}}
\def \bea {\begin{eqnarray}}
\def \eea {\end{eqnarray}}
\def \nn {\nonumber}
\def \la {\langle}
\def \ra {\rangle}
\def \rr {\raise.35ex\hbox{\small $\prime$}\kern-.17em{\mbox{\large $\imath$}}}

\def \dels {\partial\kern-.5em / \kern.5em}
\def \As {{A\kern-.5em / \kern.5em}}
\def \Ds {D\kern-.7em / \kern.5em}

\def \eps {\epsilon}
\def \m {\mu}
\def \n {\nu}

\def\frac#1#2{{#1\over #2}}

\def\Vol{\operatorname{Vol}}

\begin{document}
%%%%%%%% title page %%%%%%%%%
\thispagestyle{empty}
\begin{flushright}
{\tt YITP-13-02}
\end{flushright}

\begin{center}
\vspace{3cm} { \LARGE {Stress Tensors from Trace Anomalies in \\ 
\vskip 0.1in
Conformal Field Theories}}

\vspace{1.1cm}

Christopher P.~Herzog and Kuo-Wei Huang

\vspace{0.8cm}

{\it C.~N.~Yang Institute for Theoretical Physics \\ Stony Brook University, Stony Brook, NY  11794, USA}
%
%\vspace{0.2cm}
%{\it $\ddagger$ Department of Physics, Princeton University, Princeton, NJ 08544, USA}
%
\vspace{2cm}

\end{center}

\begin{abstract}
\noindent
Using trace anomalies, we determine
the vacuum stress tensors of arbitrary even dimensional conformal field theories in Weyl flat backgrounds. 
We demonstrate a simple relation between the Casimir energy on $\mathbb{R} \times S^{d-1}$ and the type A anomaly coefficient. 
This relation generalizes earlier results in two and four dimensions.
These field theory results for the Casimir are shown to be consistent with holographic predictions in two, four, and six dimensions.\end{abstract}

\vspace{4cm}

\noindent
January 2013

\pagebreak
\setcounter{page}{1}
%%%%%%% end of title page %%%%%%

%%%%%% table of contents %%%%%%
%\tableofcontents
%%%%%%%%%%%%%%%%%%%%%

%%%%%%%%%%%%%%%%%%%%%%%%%%%%%%%%%%%%%%%%%%%%%%%%%%%%

\subsection*{Introduction}
A conformal field theory (CFT) embedded in a curved spacetime background can be characterized by the trace anomaly coefficients of the stress tensor. Here we only consider even dimensional CFTs because there is no trace anomaly in odd dimensions.  The anomaly coefficients (or central charges) $a_d$ and $c_{dj}$ show up in the trace as follows,
\begin{equation}
\label{tracegeneral}
\langle T^\mu_\mu \rangle = \frac{1}{(4\pi)^{d/2}}\left( \sum_j c_{dj} I^{(d)}_j - (-)^{\frac{d}{2}} a_d E_d \right) \ .
\end{equation}
Here $E_d$ is the Euler density in $d$ dimensions and $ I^{(d)}_j $ are independent Weyl invariants of weight $-d$. The subscript ``$j$'' is used to index the Weyl invariants. Our convention for 
the Euler density is that
\begin{equation}
\label{Ed}
E_d = \frac{ 1}{2^{d/2}} 
%\epsilon_{\mu_1 \cdots \mu_d} \epsilon^{\nu_1 \cdots \nu_d} 
\delta_{\mu_1 \cdots \mu_d}^{\nu_1 \cdots \nu_d}
{R^{\mu_1 \mu_2}}_{\nu_1 \nu_2} \cdots
{R^{\mu_{d-1} \mu_d}}_{\nu_{d-1} \nu_d} \ . 
\end{equation}
We will not need the explicit form of the $I_j^{(d)}$ in what follows, although we will discuss their form in $d \leq 6$.

Note that we are working in a renormalization scheme where the trace anomaly is free of the so-called type D anomalies which are total derivatives that can be changed by adding local covariant but not Weyl-invariant counter-terms to the effective action.  For example, in four space-time dimensions, a $\Box R$ in the trace can be eliminated by adding an $R^2$ term to the effective action.

The constraints of conformal symmetry mean that these central charges $a_d$ and $c_{dj}$ 
determine the behavior of other correlation functions as well. In this letter, for a conformally flat background, we show how to compute $\langle T^{\mu\nu} \rangle$ in terms of $a_d$ and curvatures.
In addition to their role in determining correlation functions,
the central charges have attracted renewed interest as a way of ordering field theories under renormalization group flow.
In 2D, the classic
$c$-theorem \cite{Zamolodchikov} states that the central charge decreases through the renormalization group flow from the ultraviolet to the infrared.
In 4D,
the corresponding trace anomaly is defined by two types of central charge $c_{41}$ and $a_4$.
The conjecture that the Euler central charge $a_4$ is the analog of $c = 6a_2$ in 2D \cite{Cardy} was proven recently using dilaton fields to probe the trace anomaly \cite{Komargodski:2011vj}. The possibility of a 6D $a$-theorem was explored in \cite{Elvang:2012st}.

The properties of central charges in the 6D case are of particular interest; the (2,0) theory, which describes the low energy behavior of M5-branes in M-theory, is a 6D CFT.  From the AdS/CFT correspondence, it has been known for over a decade that quantities such as the thermal free energy \cite{Klebanov:1996un} and the central charges \cite{Henningson:1998gx} have an $N^3$ scaling for a large number $N$ of M5-branes.  However, a direct field theory computation has proven difficult.  
Any results calculated from the field theory side of the 6D CFT without referring to AdS/CFT should be interesting. Such results also provide a non-trivial check of the holographic principle.

In this letter we study the general relation between the stress tensor and the trace anomaly of a CFT in a conformally flat background. 
Our main result (\ref{mainresult}) is an expression for the vacuum stress tensor of an even dimensional CFT in a conformally flat background in terms of $a_d$ and curvatures.\footnote{%
By vacuum, we have in mind a state with no spontaneous symmetry breaking, where the expectation values of the matter fields vanish.
}
We pay special attention to the general relation between the Casimir energy (ground state energy) and $a_d$. 
Let $\epsilon_d$ be the Casimir energy on $\mathbb{R} \times S^{d-1}$.
The well known 2D CFT result is \cite{2d}
%for 2D CFTs 
%Aon $\mathbb{R}^1 \times S^1$ 
%is well-known and given by \cite{2d}
\bea
\eps_2
%|_{\mathbb{R}^1\times S^1}
=-{c\over 12 \ell} = - \frac{a_2}{2 \ell}   \ ,
\eea
where $\ell$ is the radius of $S^1$.
This result is universal for an arbitrary 2D CFT, independent of supersymmetry or other requirements.
For general $\mathbb{R} \times S^{d-1}$, we find
%\footnote{%
%Note that this relation is renormalization scheme dependent and can be altered by the presence of the so-called type D anomalies discussed in footnote 1.
%} 
\begin{equation}
\eps_d = \frac{1 \cdot 3 \cdots (d-1)}{(-2)^{d/2}} \frac{a_d}{\ell} \ .
\label{casimirresult}
\end{equation}
%and
 %{\color{blue} in our conventions, $c_{21} = 6 a_2 $.}

\subsection*{Stress Tensor and Conformal Anomaly}
We would like to determine the contribution of the anomaly to the stress tensor of a field theory in a conformally flat background.  
The general strategy we use was originally developed in \cite{A}. (See also \cite{Deser:1976yx,Page:1982fm,B,Schwimmer:2000cu} for related discussion.) The conformal (Weyl) transformation is parametrized by $\sigma(x)$
in the standard form
\bea
\bar g_{\m\n}(x)=e^{2\sigma(x)} g_{\m\n}(x) \ .
\eea
Denote the partition function as $Z[g_{\m\n}]$. The effective potential is given by
\bea
\Gamma[\bar g_{\m\n}, g_{\m\n}]= \ln Z[\bar g_{\m\n}]-\ln Z [g_{\m\n}] \ .
\eea
The expectation value of the stress tensor $\la{T^{\m\n}}\ra$ is defined by 
the variation of the effective potential with respect to the metric. 
%One could obtain
%$\la{T^{\m\n}}\ra$ when considering a conformal transformation into another manifold where the corresponding stress tensor is known. 
%
Here we consider a
conformally flat background, $\bar g_{\m\n}(x)=e^{2\sigma(x)} \eta_{\m\n}$, and we normalize the stress tensor in the flat spacetime to be zero.
The (renormalized) stress tensor is given by
\bea
\label{d}
\la{T^{\m\n}(x)}\ra={2\over\sqrt{-\bar g}} {\delta \Gamma [\bar g_{\alpha \beta}] \over \delta \bar g_{\m\n}(x)} \ ,
\eea
which implies
\bea
\label{a}
\sqrt{-\bar g} \la{T^\lambda_\lambda (x)}\ra=2\bar g_{\m\n}(x) {\delta \Gamma[\bar g_{\alpha\beta}] \over \delta \bar g_{\m\n}(x) }={\delta \Gamma[\bar g_{\alpha\beta}] \over \delta \sigma(x')} \ .
\eea
We rewrite \eqref{d} as
\bea
\label{b}
%\lefteqn{
{\delta (\sqrt{-\bar g} \la{T^{\m}_{\n}(x)}\ra)\over \delta \sigma(x')}
=
%} \nonumber \\
%&& 
2\bar g_{\lambda\rho} (x'){\delta \over \delta \bar g_{\lambda\rho}(x')}2\bar g_{\n\gamma} (x){\delta \Gamma [\bar g_{\alpha\beta}] \over \delta \bar g_{\m\gamma}(x)} \ .
\eea
Then we use the following commutative property
\bea
\Big[\bar g_{\lambda\rho} (x'){\delta \over \delta \bar g_{\lambda\rho}(x')}, \bar g_{\n\gamma} (x){\delta \over \delta \bar g_{\m\gamma}(x)}\Big]=0
\eea
to obtain the following differential scale equation
\bea
\label{final}
{\delta \sqrt{-\bar g} \la{T^{\m\n}(x)}\ra\over \delta \sigma(x')}=2{\delta \sqrt{-\bar g} \la{T^\lambda_\lambda (x')}\ra\over \delta \bar g_{\m\n}(x)} \ .
\eea
This equation determines the general relation between the stress tensor (and hence the Casimir energy) and the trace anomaly.

Next we would like to re-write the trace anomaly $\la{T^\m_\m}\ra$ in terms of a Weyl exact form, $\la{T^\m_\m}\ra= {\delta\over\delta \sigma} (\rm something)$,
so that we can factor out the sigma variation in \eqref{final} to simplify the calculation.
The integration constant is fixed to zero by taking $\langle T^{\mu\nu} \rangle = 0$ in flat space.
We use dimensional regularization and work in $n = d + \epsilon$ dimensions.
While we do not alter $E_d$ in moving away from $d$ dimensions, we will alter the form of the $I_j^{(d)}$.  Let $\lim_{n \to d} {\cal I}_j^{(d)} = I_j^{(d)}$ where ${\cal I}_j^{(d)}$ continues to satisfy the defining relation
$\delta_\sigma {\cal I}_j^{(d)} = - d \, {\cal I}_j^{(d)}$.  
We assume that in general ${\cal I}_j^{(d)}$'s exist such that
\bea
\label{E}
{\delta\over (n-d) \delta \sigma(x)} \int d^n x'\sqrt{- \bar g}{E_d}(x')&=&\sqrt{- \bar g} {E_d} \ ,\\
\label{I}
{\delta\over (n-d) \delta \sigma(x)} \int d^n x'\sqrt{- \bar g}{
{\cal I}^{(d)}_j}(x')
&=&\sqrt{- \bar g} {\cal I}^{(d)}_j \ .
\label{Injrel}
\eea

We now make a brief detour to discuss the existence of ${\cal I}_j^{(d)}$ in $d=2$, 4 and 6 \cite{Deser:1993yx} and also a general 
proof of the variation (\ref{E}).
In 2D, there are no Weyl invariants $I_j^{(2)}$ and we can ignore (\ref{Injrel}).  In 4D, we have the single Weyl invariant
$ I_1^{(4)} = C^{(n=4)}_{\mu\nu\lambda\rho} C^{(n=4) \, \mu\nu \lambda \rho}$ where $C^{(4)\mu\nu\lambda \rho}$ is the 4D Weyl tensor.  
If we define the $n$-dimensional Weyl tensor
\bea
%\lefteqn{
{C^{(n)\mu\nu}}_{\lambda\sigma}\equiv {R^{\mu \nu}}_{\lambda\sigma} 
%} \\
%&& 
- \frac{1}{n-2} \left[ 2(\delta^\mu_{[\lambda} R^\nu_{\sigma]}
+ \delta^\nu_{[\sigma} R^\mu_{\lambda]} ) +{R \, \delta^{\mu\nu}_{\lambda\sigma}\over (n-1)} \right] \ , 
%\nn
\label{Cn}
\eea
then we find ${\cal I}^{(4)}_1=C^{(n)}_{\mu\nu\lambda\rho} C^{(n) \, \mu\nu \lambda \rho}$ defined in terms of the $n$-dimensional Weyl tensor satisfies the eigenvector relation
(\ref{Injrel}).
At this point, our treatment differs somewhat from ref.\ \cite{A} where the authors vary instead $I_1^{(4)}$ with respect to $\sigma$.
While ref.\ \cite{A} allows for an additional total derivative $\Box R$ term in the trace anomaly,
in this letter we choose a renormalization scheme where
the trace anomaly takes the minimal form (\ref{tracegeneral}).
It turns out that this scheme is the one used to match holographic predictions as we will discuss shortly.
A $\Box R$ can be produced by varying $(n-4)R^2$ with respect to $\sigma$.
Such an $R^2$ term appears in the difference between ${\cal I}^{(4)}_1$ and $I^{(4)}_1$ in \cite{A}.

In 6D, there are three Weyl invariants
\bea
\label{I1}
I^{(6)}_1
&=&C^{(6)}_{\mu \nu \lambda \sigma}~C^{(6)\nu \rho\eta\lambda}~C^{(6)\mu\sigma}_{\rho~~~~~\eta} \ , \\
\label{I2}
I^{(6)}_2
&=&C^{(6)\lambda\sigma}_{\mu \nu}~C^{(6)\rho\eta}_{\lambda\sigma}~C^{(6)\mu\nu}_{\rho\eta} \ , \\
\label{I6}
I^{(6)}_3
&=&C^{(6)}_{\mu \nu \lambda \sigma}\Big(\Box\delta^{\mu}_{\rho}+4R^{\mu}_{\rho}-{6\over5} R \delta^{\mu}_{\rho}\Big) C^{(6)\rho\nu \lambda \sigma}
% \nonumber \\
%&& 
+ D_\mu J^{\mu} \ .
\eea 
To produce the ${\cal I}_j^{(6)}$ when $j=1$,2, we replace the six dimensional Weyl tensor with its 
 $n$-dimensional cousin as in the 4D case.   The variation (\ref{Injrel}) is then straightforward to show.  For $j=3$, 
 \cite{Erdmenger:1997gy} demonstrated the corresponding Weyl transformation for a linear combination
 of the three ${\cal I}^{(6)}_j$, there denoted $H$.  
  The full expression for ${\mathcal I}_3^{(6)}$ and the $n$-dimensional version of $J^{\mu}$ is not important; we refer the reader to \cite{Bastianelli:2000hi,Erdmenger:1997gy} for details.  For $d>6$, we assume the Weyl invariants can be engineered in a similar fashion;
  see \cite{Boulanger:2004zf} for the $d=8$ case.

To vary $E_d$, we write the corresponding integrated Euler density as
\begin{eqnarray}
%\lefteqn{
\int d^n x \sqrt{- \bar g} E_d = 
 \int \frac{ \left( \bigwedge_{j=1}^n dx^{\mu_j} \right) }{2^{d/2}(n-d)!}
 %} \\
% &&
%\times 
{R^{a_1 a_2}}_{\mu_1 \mu_2} \cdots
{R^{a_{d-1} a_d}}_{\mu_{d-1} \mu_d}
e^{a_{d+1}}_{\mu_{d+1}} \cdots e^{a_n}_{\mu_n} 
 \epsilon_{a_1 \cdots a_n} \ .
 % \nonumber
\end{eqnarray}
Recall that the variation of a Riemann curvature tensor with respect to the metric is a covariant derivative acting on the connection.  After integration by parts, these covariant derivatives act on either the vielbeins $e^a_\mu$ or the other Riemann tensors and hence vanish by metricity or a Bianchi identity.
Thus, in varying the integrated Euler density, we need only vary the vielbeins.  
We use the functional relation $2 \delta / \delta g^\nu_\mu = e_{(\nu}^a \delta / \delta e^a_{\mu)}$.  
  One finds
\bea
%\lefteqn{
\frac{\delta}{\delta \bar g_\mu^\nu(x)} \int d^n x' \sqrt{- \bar g} E_d =
%}\\
%&&
\frac{\sqrt{- \bar g}}{2^{\frac{d}{2}+1}} {R^{\nu_1 \nu_2}}_{\mu_1 \mu_2} \cdots
{R^{\nu_{d-1} \nu_d}}_{\mu_{d-1} \mu_d} \,  \delta^{\mu_1 \cdots \mu_d \mu}_{\nu_1 \cdots \nu_d \nu}\ . 
%\nn
\eea
From this expression, the desired relation (\ref{E}) follows after contracting with $\delta^\nu_\mu$. 

Given the variations (\ref{E}, \ref{Injrel}), we can factor out the
sigma variation in \eqref{final} to obtain\footnote{%
While we specialize to conformally flat backgrounds, under a more general conformal transformation one has
$
\la T^{\m\n}(\bar g) \ra - \la X^{\m\n} (\bar g) \ra = e^{-(d+2) \sigma} \left( \la T^{\m\n}(g) \ra - \la X^{\m\n}(g) \ra \right) 
$.
}
\bea
\lefteqn{
\la{T^{\m\n}}\ra=  \la X^{\m\n} \ra \equiv \lim_{n\to d} {1\over (n-d)} {2\over \sqrt{- \bar g} (4\pi)^{d/2}}
 }  \\
&&
\times 
{\delta\over \delta \bar g_{\m\n}(x)}\int d^n x' \sqrt{-\bar g} \left({\sum_j c_{dj} {\cal I}^{(n)}_j - (-)^{\frac{d}{2}} a_d E_d} \right) . 
\nn
\eea
Comparing with (\ref{d}), we see that the effective action must  contain terms proportional to $\la T^\m_\m \ra$.  Indeed, these are precisely the counter terms that must be added to regularize divergences coming from placing the CFT in a curved space time \cite{BD}.
We next perform the metric variation for a conformally flat background.
The metric variation of the Weyl tensors ${\cal I}_j^{(d)}$ vanishes for conformally flat backgrounds because the ${\cal I}_j^{(d)}$ are all at least quadratic in the $n$-dimensional Weyl tensor. (Conformal flatness is used only after working out the metric variation.) Thus the stress tensor in a conformally flat background may be obtained by varying only the Euler density:
\bea
\label{mainresult}
%\lefteqn{
\langle T^\mu_\nu \rangle = - \frac{a_d}{(-8 \pi)^{d/2}}
%} 
% \\
%&& \times
 \lim_{n \to d} \frac{1}{n-d} 
 {R^{\nu_1 \nu_2}}_{\mu_1 \mu_2} \cdots
{R^{\nu_{d-1} \nu_d}}_{\mu_{d-1} \mu_d} \,  \delta^{\mu_1 \cdots \mu_d \mu}_{\nu_1 \cdots \nu_d \nu}\, .
% \nn
\eea
Note that in a conformally flat background, employing (\ref{Cn}), the Riemann curvature can be expressed purely in terms of the Ricci tensor and Ricci scalar:
\[
{R^{\nu_1 \nu_2}}_{\mu_1 \mu_2} = \frac{1}{n-2} \left[ 2( \delta^{\nu_1}_{[\mu_1} R^{\nu_2}_{\mu_2]} + \delta^{\nu_2}_{[\mu_2} R^{\nu_1}_{\mu_1]}) - \frac{ R \, \delta^{\nu_1 \nu_2}_{\mu_1 \mu_2} }{n-1} \right] \ .
\]
Contracting a $\delta^{\nu_j}_{\mu_j}$ with the antisymmetrized Kronecker delta $ \delta^{\mu_1 \cdots \mu_d \mu}_{\nu_1 \cdots \nu_d \nu}$ eliminates the factor of $(n-d)$ in (\ref{mainresult}).

In 2D and 4D, we can use (\ref{mainresult}) to recover results of \cite{A}. In 2D, the right hand side of $\la T^\m_\n \ra$ is proportional
to $R^\m_\n - \frac{1}{2} R \delta^\m_\n$ which vanishes in 2D.  Thus we first must expand the Einstein tensor in terms of the Weyl factor $\sigma$ where $g_{\m\n} = e^{2 \sigma} \eta_{\m\n}$ before taking the $n \to 2$ limit.   The result is \cite{A}
\bea
\la{T^{\m\n}}\ra&=&{a_2\over 2\pi} \left( \sigma^{,\m;\n} +\sigma^{,\m}\sigma^{,\n} -g^{\m\n}
\left( 
%g^{\lambda \rho}
{\sigma_{,\lambda}}^{;\lambda}
+\sigma_{,\lambda}\sigma^{,\lambda}\right)
\right) \ .
\eea
In 4D, we obtain 
\bea
\la{T^{\m\n}}\ra={-a_4\over(4\pi)^2} \Big[g^{\m\n} \Big({R^2\over 2}-R^2_{\lambda \rho} \Big)+2R^{\m\lambda}R^{\n}_\lambda-{4\over3}RR^{\m\n}\Big] \ .
\label{fourDresult}
\eea
In 6D, we obtain (to our knowledge) a new result
\bea
\label{mainresult6D}
%&&
\la{T^{\m\n}}\ra &=&-{a_6 \over (4\pi)^3} \left[\frac{3}{2} R_{\lambda}^\m R_{\sigma}^\n R^{\lambda \sigma} -
\frac{3}{4} R^{\m\n} R^{\lambda}_{\sigma} R_{\lambda}^{\sigma}
% \nn\\
%&&
 -\frac{1}{2} g^{\m\n} R_\lambda^\sigma R^{\lambda}_{\rho} R^{\rho}_{\sigma} \right.
\nn \\
&&
\left.
 -
\frac{21}{20} R^{\m \lambda} R^\n_{\lambda} R + \frac{21}{40} g^{\m\n} R_{\lambda}^{\sigma}R^{\lambda}_{\sigma} R
%\nn\\
%&& 
+ \frac{39}{100} R^{\m\n}R^2- \frac{1}{10} g^{\m\n} R^3\right] \ .
\eea
As we work in Weyl flat backgrounds, there is no contribution from B type anomalies.
These $\la{T^{\m\n}}\ra$ are covariantly conserved, as they must be since they were derived from a variational principle.

\subsection*{Casimir Energy and Central Charge}
We would like to relate $a_d$ to the Casimir energy 
\bea
\label{epsdef}
\eps_d=  \int_{ S^{d-1}}  \la{T^{00}}\ra \operatorname{vol}(S^{d-1}),
\eea
on $\mathbb{R} \times S^{d-1}$.  
In preparation, 
let us calculate $E_d$ for the sphere $S^d$.
For $S^d$ with radius $\ell$, the Riemann tensor is ${R^{\nu_1 \nu_2}}_{\mu_1 \mu_2} = \delta^{\nu_1 \nu_2}_{\mu_1 \mu_2}/ \ell^2$.  
It follows from (\ref{Ed}) that 
%\begin{equation}
$E_d = \frac{d!}{\ell^d}.$ %\ .
We conclude that the trace of the vacuum stress tensor on $S^d$ takes the form
\begin{equation}
\langle T^\mu_\mu \rangle = - \frac{a_d \, d!}{(-4 \pi \ell^2)^{d/2}} \ .
\end{equation}
Let us now calculate $\langle T^\mu_\nu \rangle$ for $S^1 \times S^{d-1}$. 
The Riemann tensor on $S^1 \times S^{d-1}$ is zero whenever it has a leg in the $S^1$ direction and looks like the corresponding Riemann tensor for $S^{d-1}$ in the other directions.  We can write ${R^{i_1 i_2}}_{j_1 j_2} =
%2 \delta^{j_1}_{[i_1} \delta^{j_2}_{i_2]}
\delta_{j_1 j_2}^{i_1 i_2}
 / \ell^2$, where $i$ and $j$ index the $S^{d-1}$.  
The computation of $\langle T^0_0 \rangle$ and $\langle T^i_j \rangle$ proceeds along similar lines to the computation of $E_d$:
\bea
\langle T^0_0 \rangle = - \frac{a_d(d-1)!}{(-4 \pi \ell^2)^{d/2}} \ ,~~\langle T^i_j \rangle = \frac{a_d(d-2)!}{(-4 \pi \ell^2)^{d/2}} \delta^i_j \ .
\eea
Note that $\langle T^\mu_\nu \rangle$ is traceless, consistent with a result of \cite{B}. Using the definition (\ref{epsdef}), we compute the Casimir energy $\eps_d$. 
We find that (for $d$ even)
\begin{equation}
\epsilon_d = \frac{a_d(d-1)!}{(-4 \pi \ell^2)^{d/2}} \Vol(S^{d-1}) = \frac{1 \cdot 3 \cdots (d-1)}{(-2)^{d/2}} \frac{a_d}{\ell} \ .
\end{equation}  
In 2D, 4D and 6D, the ratios between the Casimir energy and $a_d$ are 
$-{1\over 2\ell}$, $3\over 4\ell$ and $-{15\over 8\ell}$, respectively.

\subsection*{Holography and Discussion} 
In this section, we would like to use the AdS/CFT correspondence to check our relation between $\epsilon_d$ and $a_d$ for $d=2$, 4 and 6.
For CFTs with a dual anti-de Sitter space description, the stress-tensor can be calculated from a
classical gravity computation \cite{Balasubramanian:1999re, Emparan:1999pm,deHaro:2000xn}.
The Euclidean gravity action is taken to be
%\begin{equation}
\bea
S &=& S_{\rm bulk} + S_{\rm surf} + S_{\rm ct} \ , \\ 
S_{\rm bulk} &=& -\frac{1}{2 \kappa^2} \int_{\cal M} d^{d+1}x \sqrt{G} \left( {\cal R} + \frac{d(d-1)}{L^2} \right), \nn\\
S_{\rm surf} &=& - \frac{1}{ \kappa^2} \int _{\cal \partial M}d^d x \sqrt{g} K , \nn\\
S_{\rm ct} &=& \frac{1}{2 \kappa^2} \int _{\cal \partial M}d^d x \sqrt{g} \Bigl[\frac{2(d-1)}{L} + \frac{L}{d-2} {R} +\nn\\
&& \frac{L^3}{(d-4)(d-2)^2}
\left( {R}^{\m\n} {R}_{\m\n} - \frac{d}{4 (d-1)} {R}^2 \right) + \ldots \Bigr] .\nn
\eea
%\end{eqnarray*}
The Ricci tensor ${R}_{\m\n}$ is computed with respect to the boundary metric $g_{\m\n}$ while ${\cal R}$ is the Ricci Scalar computed from the bulk metric $G_{ab}$.
The object $K_{\m\n}$ is the extrinsic curvature of the boundary $\partial \cal M$.
The counter-terms $S_{\rm ct}$ render $S$ finite, and we keep only as many as we need.
The metrics with $S^{d-1} \times S^1$ conformal boundary,
\begin{equation}
ds^2 = L^2( \cosh^2 r \, dt^2 + dr^2 + \sinh^2 r \, d\Omega_{d-1}) \ ,
\end{equation}
and $S^{d}$ boundary,
\begin{equation}
ds^2 = L^2 ( dr^2 + \sinh^2 r \, d\Omega_{d} ) \ ,
\end{equation}
satisfy the bulk Einstein equations.
Note that the $S^{d-1}$ and $S^d$ spheres have radius $\ell = \frac{L}{2} e^{r_0}$ at some large reference $r_0$ while we take the $S^1$ to have circumference $\beta$ (hence the range of $t$ is $0 < t < \beta / \ell$). We compute the stress tensor from the on-shell value of the gravity action using (\ref{d}),
making the identification $\Gamma = -S$ and using the boundary value of the metric in place of $\bar g_{\mu\nu}$.
%The following values were computed in 
One has \cite{Emparan:1999pm}:
\begin{equation*}
\renewcommand\arraystretch{1.2}
\begin{array}{|c|c|c|}
\hline
d & \Gamma_{S^d} & \Gamma_{S^1 \times S^{d-1}} \\
\hline
2 &\frac{4 \pi L}{\kappa^2} \log \ell & \frac{\pi \beta L}{\kappa^2 \ell}\\
\hline
4 & - \frac{4 \pi^2 L^3}{\kappa^2} \log \ell & - \frac{3\pi^2 \beta L^3}{4\kappa^2 \ell}\\
\hline
6 & \frac{2 \pi^3 L^5}{\kappa^2} \log \ell & \frac{5 \pi^3 \beta L^5}{16 \kappa^2 \ell}
\\
\hline
\end{array}
\end{equation*}
We include only the leading log term of $\Gamma_{S^d}$. 
From (\ref{d}), it follows that
%\begin{eqnarray}
$\langle T^0_0 \rangle  \Vol(S^{d-1}) = \partial_\beta \Gamma_{S^1 \times S^{d-1}} $ and
 $\langle T^\mu_\mu \rangle   \Vol(S^d) = \partial_\ell \Gamma_{S^d} $
For a conformally flat manifold, we have from (\ref{tracegeneral}) that $\langle T^\mu_\mu \rangle = - a_d (-4\pi)^{-d/2} E_d$ which allows us to calculate $a_d$ from $\langle T^\m_\m \rangle$ \cite{Henningson:1998gx}.
Defining the Casimir energy with respect to a time $\tilde t = \ell t$ whose range is the standard $0 < \tilde t < \beta$,
we can deduce from (\ref{epsdef}) that $\epsilon_d = - \partial_\beta \Gamma_{S^1 \times S^{d-1}}$ (see also \cite{Awad:2000aj}).
We have a table:
\begin{equation*}
\renewcommand\arraystretch{1.2}
\begin{array}{|c|c|c|c|c|c|c|}
\hline
& \langle T^0_0 \rangle & \epsilon_d & & \langle T^\mu_\mu \rangle &E_d & a_d\\
\hline
S^1 \times S^1 & \frac{L}{2 \kappa^2 \ell^2} & -\frac{\pi L}{\kappa^2 \ell} & S^2 & \frac{L}{\kappa^2 \ell^2} & \frac{2}{\ell^2} & \frac{2 \pi L}{\kappa^2} \\
\hline
S^1 \times S^3 & -\frac{3 L^3}{8 \kappa^2 \ell^4} & \frac{3 \pi^2 L^3}{4 \kappa^2 \ell} & S^4 & - \frac{3 L^3}{2 \kappa^2 \ell^4} & \frac{24}{\ell^4} & \frac{\pi^2 L^3}{\kappa^2}\\
\hline
S^1 \times S^5 & \frac{5 L^5}{16 \kappa^2 \ell^6} & -\frac{5 \pi^3 L^5}{16 \kappa^2 \ell} & S^6 & \frac{15 L^5}{8 \kappa^2 \ell^6}& \frac{720}{\ell^6}& \frac{\pi^3 L^5}{6 \kappa^2} \\
\hline
\end{array}
\end{equation*}
Comparing the $\eps_d$ and $a_d$ columns, we can confirm the results from earlier in this paper, namely that\footnote{%
These results indicate that any so-called type D anomalies present in the holographic renormalization scheme do not
afffect the relation between $a_d$ and $\epsilon_d$ determined in a scheme where the type D anomalies are absent.
}
\bea
\eps_2 = - \frac{a_2}{2 \ell} \ ; \; \; \;
\eps_4 = \frac{3 a_4}{4 \ell} \ ; \; \; \;
\eps_6 = - \frac{15 a_6}{8 \ell} .
\eea

In the 4D case, such a gravity model arises in type IIB string theory by placing a stack of $N$ D3-branes at the tip of a 6D Calabi-Yau cone. In this case, we can make the further identification \cite{Aharony:1999ti,Henningson:1998gx}: 
%\bea
$a_4 = \frac{N^2}{4} \frac{\Vol(S^5)}{\Vol(SE_5)}$
%\eea
where $SE_5$ is the 5D base of the cone. These constructions are dual to 4D quiver gauge theories with ${\mathcal N}=1$ supersymmetry. In 6D, such a gravity model arises in M-theory by placing a stack of $N$ M5-branes in flat space. In this case, we can make the further identification \cite{Henningson:1998gx,Bastianelli:2000hi} (see also \cite{Maxfield:2012aw}):
%\bea
$a_6 = \frac{N^3}{9}.$
%\eea
The dual field theory is believed to be the non-abelian (2,0)-theory.

We would like to comment briefly on the Casimir energy calculated in the weak coupling limit.\footnote{%
We thank J.~Minahan for discussions on this issue.
} 
In typical regularization schemes, for example zeta-function regularization, the Casimir energy will not be related to the conformal anomaly via (\ref{casimirresult}) because of the presence of total derivative terms (D type anomalies) in the trace of the stress tensor.  
For a conformally coupled scalar in 4D, ref.\ \cite{BD} tells us $a_4 = 1/360$.  
Our result (\ref{casimirresult}) would imply then that $\epsilon_4 = 1/480L$, 
but naive zeta-function regularization yields instead $\epsilon_4 = 1/240L$.   
The discrepancy can be resolved either by including a $\Box R$ term in the trace, thus changing (\ref{casimirresult}) \cite{B}, 
or by adding an $R^2$ counter-term to the effective action, thereby changing $\epsilon_4$.
Amusingly in zeta-function regularization, the effect of the total derivative terms on $\epsilon_4$ 
cancels for the full ${\mathcal N}=4$ SYM multiplet, and the weak coupling results for
$\epsilon_4$ and $a_4$ are related via (\ref{casimirresult}) \cite{Ford:1976fn, Marino:2011nm}. 
In contrast, for the (2,0) multiplet in 6D, the total derivative terms do not cancel \cite{Bastianelli:2000hi}.  The resulting discrepancy
\cite{Gibbons:2005jd} in the relation between $a_6$ and $\epsilon_6$ can presumably be cured either by adding counter-terms
to the effective action to eliminate the total derivatives or by improving (\ref{casimirresult}) to include the effect of these derivatives.
Generalizing our results to include the contribution of D type anomalies to the stress tensor would allow a more
straightforward comparison of weak coupling Casimir energies obtained via zeta-function regularization and the conformal anomaly $a_d$.  
We leave such a project for the future.
%The claim in this paper is that one finds ratios $3\over 4$ and $- {15\over 8}$ in 4D and 6D, respectively, in a scheme where the effect of the total derivative terms in the trace vanish (no type D anomaly); the holographic computation uses precisely such a scheme.

%Given our result (\ref{casimirresult}), 
%there is a mismatch between the Casimir energy computed in \cite{Gibbons:2005jd} and the conformal anomaly computed in
%\cite{Bastianelli:2000hi}.
%%regarding the Casimir energy of the 6D (2,0) theory in \cite{Awad:2000aj} and \cite{Gibbons:2005jd}.
%%Using the central charge given in \cite{Bastianelli:2000hi}, our relation $\eps_6=-{15\over 8\ell} a_6$ does not match the Casimir energy obtained in \cite{Gibbons:2005jd}. 
%Here is our interpretation. Unlike the case of $\rm{N=4~ SYM}$, where the coefficients for Type D anomaly can be shown to be zero \cite{BD},  the trace anomaly of 6D (2,0) theory would get scheme dependent contributions from total derivative terms. Moreover, we conjecture that the zeta function regularization corresponds to a scheme where these total derivative terms are present for 6D (2,0) theory. The claim in this paper is that one would find ratios $3\over 4$ and $- {15\over 8}$ in 4D and 6D, respectively, in a scheme where these total derivative terms vanish (no type D anomaly) and the holographic computation uses precisely such a scheme.

There are two other obvious calculations for future study: i) Determine how $\la{T^{\m\n}}\ra$ transforms in non-conformally flat backgrounds. %not just the one from flat space considered in this letter.%
Such transformations would involve the type B anomalies.
ii) %Regarding our holographic check, we focused on two special backgrounds $S^d$ and $S^{d-1} \times S^1$. In principle, one should be able to
Check the full 6D stress tensor (\ref {mainresult6D}) for any conformally flat background by the holographic method. A 4D check of (\ref{fourDresult}) was performed in \cite{de Haro:2000xn}. %To our knowledge, the analogous 6D computation has not yet been done.
%iii) Perform field theory computations of the Casimir energy for (2,0) theory. We expect to reproduce $\eps_6 = -15 a_6 / 8 \ell$. %\CH{worried (iii) is trivial}
%
%iii) Generalize our results to include the contribution of D-type anomalies to the stress tensor.  
%Such a generalization should allow a more straightforward comparison of weak coupling Casimir energies obtained via zeta-function regularization and the conformal anomaly $a_d$.  
%In the weak coupling limit, use the zeta regularization method to obtain type D anomaly contributions to Casimir energy and match
%the results calculated from the trace anomaly approach.
%
%
\\\\
\noindent{\bf Acknowledgments}:
We would like to thank N.~Bobev, Z.~Komargodski, J.~Minahan, M.~Ro\v{c}ek, Y.~Nakayama, P.~van~Nieuwenhuizen, A.~Schwimmer, and R.~Vaz for discussion.
The Mathematica
packages \cite{math} were useful for checking our results.
This work was supported in part by the National Science Foundation under Grants No. PHY-0844827 and PHY-0756966.  C.~H. also thanks the Sloan Foundation for partial support.
%%%%%%%%%%%%%%%%%%%%%%%%%%%%%%%%%%%%%%%%%%%%%%%%%%%%%%%%%%%%%%%%%%%%%%%%%%%%%%%%%%

\end{document}